\documentclass[a4paper,fleqn]{cas-dc}

\usepackage[numbers]{natbib}
\usepackage{xurl}
\usepackage{xcolor}
\usepackage{soul}
\usepackage{verbatim}

\def\tsc#1{\csdef{#1}{\textsc{\lowercase{#1}}\xspace}}
\tsc{WGM}
\tsc{QE}
\tsc{EP}
\tsc{PMS}
\tsc{BEC}
\tsc{DE}

\begin{document}
\let\WriteBookmarks\relax
\def\floatpagepagefraction{1}
\def\textpagefraction{.001}
\shorttitle{A Serverless Cloud Integration For Quantum Computing}
\shortauthors{M.Grossi, L.Crippa, A.Aita et~al.}

\title [mode = title]{A Serverless Cloud Integration For Quantum Computing}                      




\author[ibm]{M. Grossi}[orcid=0000-0003-1718-1314]
\credit{Conceptualization, Methodology, Software, Writing - Original Draft, Review \& Editing}

\author[ibm,parma]{L. Crippa}[orcid=0000-0003-1492-9542]
\credit{Conceptualization, Methodology, Software, Writing - Original Draft, Review \& Editing}

\author[ibm]{A. Aita}
\credit{Conceptualization, Methodology, Software, Writing - Original Draft, Review \& Editing}

\author[ibm]{G. Bartoli}
\credit{Methodology, Software Design, Writing, Review \& Editing}

\author[ibm]{V. Sammarco}
\credit{Methodology, Software Design, Writing, Review \& Editing}

\author[ibm]{E. Picca}
\credit{Software and Formal analysis, Writing, Review \& Editing}

\author[ibm]{N. Said}
\credit{Software and Formal analysis, Writing, Review \& Editing}

\author[cic]{F. Tramonto}[orcid=0000-0003-4540-5004]
\credit{Software and Formal analysis, Writing, Review \& Editing}

\author[ibm]{F. Mattei}
\credit{Supervision and validation}

\address[ibm]{IBM Italia S.p.A., Circonvallazione Idroscalo, 20090 Segrate (MI), Italy}

\address[parma]{Università di Parma, Dipartimento di Fisica, Parco Area delle Scienze, 7/A, 43124 Parma, Italy}

\address[cic]{IBM Client Innovation Center Srl, Via Lombardia 2/A, 20068 Peschiera Borromeo (MI), Italy}

\begin{abstract}
Starting from the idea of Quantum Computing which is a concept that dates back to 80s, we come to the present day where we can perform calculations on real quantum computers. This sudden development of technology opens up new scenarios that quickly lead to the desire and the real possibility of integrating this technology into current software architectures. 
The usage of frameworks that allow computation to be performed directly on quantum hardware poses a series of challenges.\\
This document describes a an architectural framework that addresses the problems of integrating an API exposed Quantum provider
in an existing Enterprise architecture and it provides a minimum viable product (MVP) solution that really merges classical quantum computers on a basic scenario with reusable code on GitHub repository. The solution leverages a web-based frontend where user can build and select applications/use cases and simply execute it without any further complication. Every triggered run leverages on multiple backend options, that include a scheduler managing the queuing mechanism to correctly schedule jobs and final results retrieval. The proposed solution uses the up-to-date cloud native technologies (e.g. Cloud Functions, Containers, Microservices) and serves as a general framework to develop multiple applications on the same infrastructure.
\end{abstract}




\maketitle

\section{Introduction}

The design of a software on distributed system is based on essential pillars such as modularity, openness and reuse of components. Typically, the application is divided into logical layers allowing targeted interventions on decoupled elements~\cite{coulouris2005distributed}. However, the usage of frameworks that allow computation to be performed on quantum hardware poses a series of challenges that will be extensively discussed in the next section.

The goal of this project is to overcome these constraints and develop a software architecture that can be reused as a design pattern whenever dealing with similar problems. The result is a system able to receive requests from the user, send them to a quantum computer and receive back the result by assuring the ordering and coherence of events  as well as the right format~\cite{lamport2019time}. At the same time, the system must maintain the characteristic of openness: it would be possible to plug and unplug new functionalities. A first attempt to create a modular design to integrate a quantum computer API services within a three tier application was made by some of the authors in~\cite{qflask}. In that case, the problem of managing an asynchronous service was not addressed and the focus was on a different technology for the backend integration.

Given this requirement, we decided to adopt a configuration driven architecture where it is possible to add/remove components with minimal effort and without impacting the other parts of the system, which are loosely decoupled~\cite{gomaa2004software}. On the other hand, being notified when the computation is done has been achieved by exploiting an Event Streams queue hosted on IBM Cloud~\cite{ibm:evstr}, built on top of Apache Kafka~\cite{thein2014apache}. The usage of a streaming application that reacts when a message is published on the channel, according to a pu\-blish/sub\-scribe pattern~\cite{eugster2003many}, is the proposed solution towards a real-time system. However, since Event Streams promotes parallelism as the number of consumers listening on a specific topic, we had to develop a user-specific assignment as a secure procedure to manage multi topic requests.
The article is organized in the following: in Sec.\,\ref{sec:soa} we describe the state of the art related to the integration of a new technology like the quantum computer in the current IT scenario, focusing on the effective challenges and feasible solutions. 
In Sec.\,\ref{sec:prop_arch} we elaborate on the proposed architecture from a general perspective defining the data flow and what are the current requirements and limitations in order to provide a possible best practice in implementing a real quantum computing based application. In Sec.\,\ref{sec:prop_frame} we describe in details each component of the proposed architecture, not only from a generic technological component but also suggesting specific items available on the market. We conclude this paper with a schematic reconstruction of the proposed solution with a remark about motivations, the technology and the methodology adopted in Sec.\,\ref{sec:concl}.

\section{State of Art}
\label{sec:soa}
The idea of Quantum Computing is a concept that dates back to 80s, thanks to the work of Benioff~\cite{Benioff} that developed a model for a quantum mechanical touring machine. Few years later, Feynman in his paper~\cite{Feynman:1981tf} proposed the idea that to simulate quantum systems one should use a computer responding to the laws of quantum mechanics. Feynman speculated that the synergistic usage of quantum super-position and entanglement in computation may enable the design of computing devices showing a high degree of parallelism, which grows with the size of the device itself, in exponential way.
While, at first, the physical build-up of such a system was something out of reach, the theoretical work on the field has started and in the 90s the most famous algorithms like the Shor's~\cite{365700} and Grover's~\cite{10.1145/237814.237866} were already available, and we understood that quantum computers could be used for more tasks than quantum simulations only.
In 1996 Di Vincenzo proposed a set of minimal ‘criteria’ for a physical system to be considered a Quantum Computer~\cite{DiVincenzo_2000}.
The unit information of a quantum computer is the qubit, and there are several physical systems that can act as qubits. During late 90s, a lot of candidates for quantum bits started to show up, and today we can count a lot of different promising technology, some of which are superconductors~\cite{Kjaergaard_2020}, ion traps~\cite{doi:10.1063/1.5088164}, photonic~\cite{Arrazola_2021}, each of which has some pros and cons at this time in terms of coherence time, working temperature, easy-control, scalability.
The systems we know today are still prototypical, in a sense that they are very good to learn and test the technology, but they are still not able to provide a real advantage in terms of computation time with respect to classical computers, even if a claim for a ‘Quantum Supremacy’ proof has been called out by Google in 2019~\cite{google}.
The paradigm for the quantum computing architecture of today - that will probably stay for years - is that of a service that is available as a Cloud service. The first 5 qubits available publicly on the Cloud were released by IBM in 2016~\cite{ibmq:exp16} and now the ‘IBM Quantum Experience’ has several processors available in the Cloud to be used for free. 
The performance of a Quantum computer are clearly related to the number of qubits, but also to other factors like the noise, the errors and so on. IBM Researchers proposed a way to measure this performance named Quantum Volume~\cite{PhysRevA.100.032328}. Quantum Error Correction, that started in the 90s, and control software are presently very important ways to improve the performance of existing systems. 
Current cloud architecture are focused only on quantum oriented experience where the user can create circuits to be run, can select a device on which to run them and then can send the job to the service. The job will be queued together with others and run when the device is available. The whole path, however, is not connected with ‘classical’ computation, no management of a quantum algorithm definition with respect to real input is taken into account.
The inclusion of a quantum algorithm execution, from data encoding to the return of processed data for a classical computation flux needs to be addressed.
The aim of this paper is indeed to propose a reference architectural framework able to bridge the gap between the classical and quantum computation for real problems.
At the time of writing there are no other proposals in this sense. We provide then a general reference framework as well as a working minimum valuable product with a specific technology adoption in the context of hybrid cloud.

\subsection{Challenges}
\label{sec:chall}

With the rising of quantum technologies (like IBM Quantum system~\cite{ibm:qexp}), enterprises and researchers will be likely to use this kind of technology almost on a daily basis in a relatively near future. Thinking about a real scenario, an Enterprise would probably be oriented in progressively integrate quantum technologies in their up and running architecture to support and improve existing workloads rather than entirely replace existing workloads with quantum technologies.
So, with this assumption, we developed an architectural framework that addresses the problems of integrating an API exposed Quantum provider, like the IBM Quantum system, in an existing Enterprise architecture. There are two major challenges in this: the first one is related to the technological and theoretical knowledge requirements in adopting a quantum provider: currently the only way to integrate a regular workload (e.g. a web application or a scheduled batch) is to use Qiskit SDK~\cite{ibm:qdev} or the Rigetti SDK~\cite{rig:qdev}, that represent, at this time of writing, the only two real quantum computer full stack systems. In this study, we focused on Qiskit SDK. Nevertheless, the same limitations and challenges that will be presented in this chapter, apply to any Qiskit like API indiscriminately. Qiskit is an SDK currently available only for Python, this means that, without the proper decoupling logic, the only workloads integrable with IBM Quantum system would be Python workloads. Moreover, even in a scenario in which a Python workload needs to be integrated with IBM Quantum system, in order to use Qiskit, a developer must know the logic of Quantum Circuit~\cite{ibm:circ} composition. In an Enterprise scenario, where a certain amount of effort is required to change the developers team composition once it has been defined (from different perspectives: timing, cost, logistics), using Qiskit can become harsh~\cite{VOLKOFF2004279}. In general, we are talking about an accessibility problem, that is common also in the integration of a standard workload with HPC environments~\cite{WONG20131333}. In fact, the second type of challenges is related to the similarities between IBM Quantum system and an HPC environment. These two types of computational services share some common features such as: the complexity of the calculations and therefore the large process times, the concurrent access to the hardware recourse from different client systems that introduces a workload submission management and a scalability problem. We decided to develop this framework on the cloud (IBM Cloud) for the extended number of services that an enterprise grade cloud provider has in terms of data storage, hosting and middleware technologies and message queues. The enterprise grade support and SLA's of the major cloud providers allow the team to focus only on the component design and development rather than infrastructural and availability problems.

\begin{figure*}[ht]
\centering
\includegraphics[width=.9\textwidth]{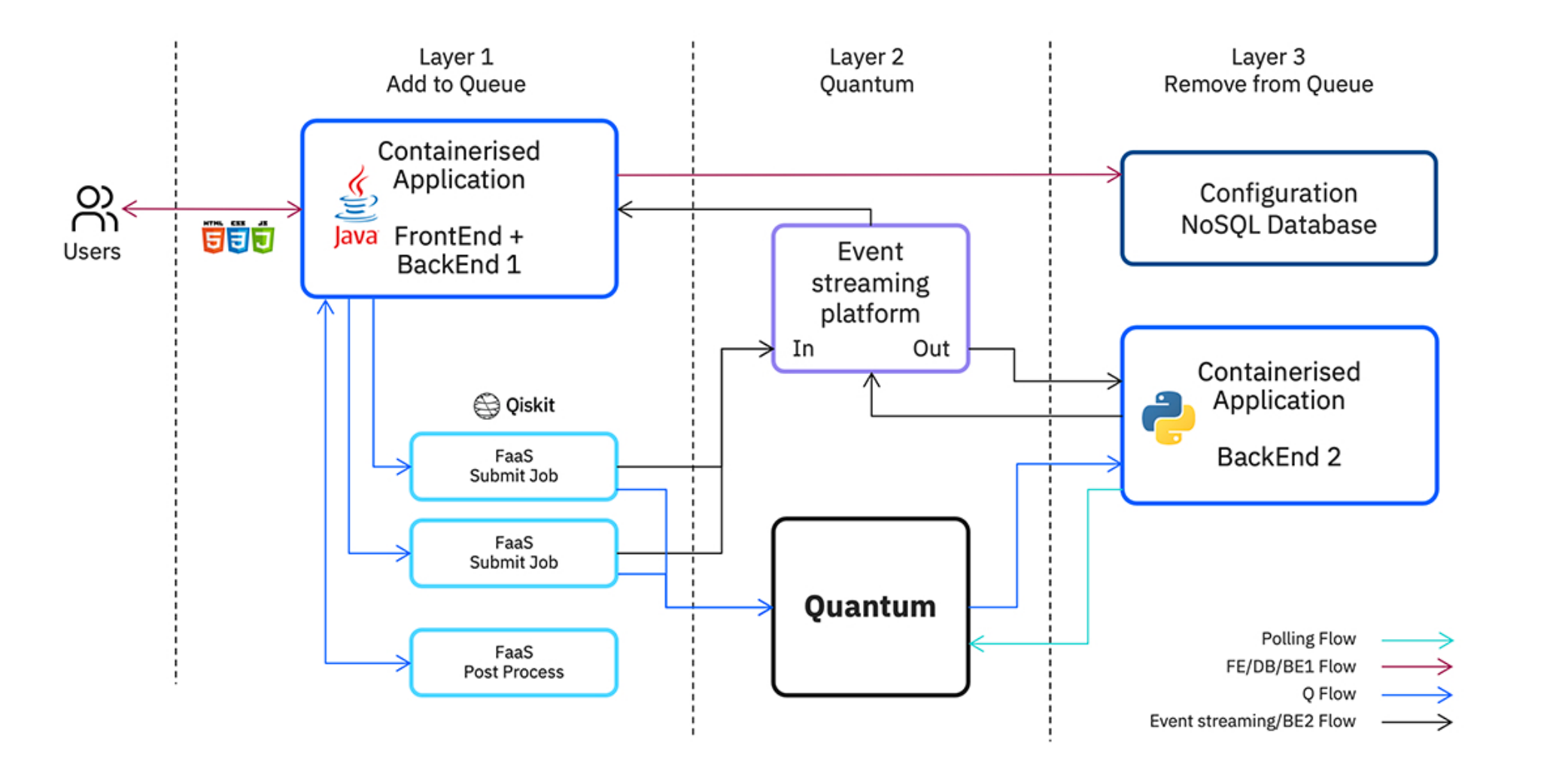}
\caption{Architectural overview diagram and flow.}
\label{img:aod}
\end{figure*}

\section{Proposed architecture, requirements and data flow}
\label{sec:prop_arch}
The framework developed aims to solve the two major challenges of accessibility and workload management. The accessibility challenge is tackled using a serverless FaaS (Functions as a Service) technology~\cite{fox_status_2017} that exposes a language agnostic HTTP interface to receive input data. This kind of technology offers a finer grain computational unit and is designed to execute small amounts of code in short time when a specific action is performed (a trigger). This can be done both by deploying a single packaged and executable software specifying the desired runtime (e.g. a .jar file with a Java runtime) or by deploying a custom Docker container~\cite{openwhisk}. To achieve accessibility, we designed a Docker container with Python as base image\footnote{https://hub.docker.com/r/ibmfunctions/action-python-v3.7}  extended with Qiskit, the open-source SDK developed by IBM and by the Qiskit Community to program quantum computers, as quantum integration framework that will expose a single HTTP API to receive raw data and transform it into a Quantum Circuit. The Framework provides a single Docker container (hence a single function with a single HTTP endpoint) for each specific algorithm that will be implemented. This is the decoupling logic that the Framework implements, hence it requires the availability of such technology in the environment chosen for deployment. In this way, a dedicated team can work only on the FaaS platform to expose multiple quantum algorithms through an HTTP API and a different system administration team can invoke them simply calling the designated endpoint, without worrying about the implementation of the algorithm itself. It is important to note that FaaS technology addresses also another architectural requirement: the scalability. Since the Framework operates as a facilitator for accessing the IBM Quantum system, its components will be used by multiple external systems; this implies that it needs to be highly scalable to avoid critical bottlenecks in the job submission phase. Here comes another analogy with the HPC architectures: the job scheduling system. This is crucial for the overall performances of the whole framework~\cite{REUTHER201876}. Due to the scalability offered by the FaaS technology,  in order to face an increasing amount of requests, the Framework gives the possibility for many external systems to be integrated with it without loss of performances, handling Quantum Circuit composition and job submission. This is a further requirement of the FaaS component to achieve scalability. Despite the good scalability level in the phase of job submission, to be effectively scalable, the Framework needs another component to increase the scalability in the phase of results retrieval. The adopted component is a queue with a kafka interface. This queue component is used both by the FaaS element, that writes a submission report on a specific topic, containing all the information needed to retrieve the output of the job together with the identifier of the client that submitted it, and by the results collector component (a Python application further referenced as Backend 2) to retrieve the job results and send them as a message on another specific topic of the queue. An important prerequisite for the queue component is the high throughput, since it needs to handle all the messages received by the FaaS component during the job submission phase. As per the user interface component (a Java application further referenced as Backend 1) and results retriever component (Python application)  the only architectural requirement is the availability of a general Java and Python runtimes to run the code. Since the code is ready to be containerized, also a standard container runtime (like Docker or Podman) and a container registry would work perfectly fine. In the context of the MVP we chose Cloud Foundry as runtime environment since it offers auto-configuration capabilities and handy administration interfaces that reduce the time and the effort needed for set it up.
In Fig. \ref{img:aod} we show the proposed overall Architecture overview diagram and flow, as a first hybrid quantum-classical approach reflected to three different layers. 
In the following paragraph we describe a specific flow according to the first simple web application built: the representation of multiple 2-characters emoticons using the superposition property of quantum computing.

A user opens the user interface (UI) in a web browser and inserts two emoticons to be visualized in superposition. A “submit button” triggers the Layer 1 to add a job in the queue and it sets itself as a listener of the outcomes. Like a scheduler, backend system engages the Event Streams and the IBM Quantum computer to submit the job in the queue on Layer 2. 
Once the quantum computer has provided the computation outcome, IBM Cloud Foundry backend in Layer 3 is engaged and the results are processed.
Finally, the UI gets unlocked and the results are displayed. 

\section{Framework components}
\label{sec:prop_frame}

\subsection{FrontEnd}

Generally, the proposed framework User Interface (UI) leverages on a set of JavaScript functions, asynchronous Ajax calls, and a well-defined data structure to get inputs and to read outputs. All the graphic comparts of HTML and CSS, together with the “general” JavaScript functions used to build the UI itself, can be modelled to fit customer's need and templates: it can be implemented in a “vanilla” flavours such as the one included prototype, or using any kind of ready to use a template (i.e., Bootstrap, React, etc.). 
The FrontEnd layer has the role of input collector and result displayer: it uses a set of client-side functions to perform the main integration tasks. The UI is built using the Carbon Design System components and functions, leveraging on the mobile-first approach, and ensuring cross-browser and cross-device compatibility. 
The FrontEnd is built as a general UI containing all applications developed under the proposed framework. However, it can be adapted to the user's need by changing the structure and graphics of the pages, leaving unaltered the JavaScript functions while leveraging on the proposed framework to communicate with the BackEnd 1. 
When a user interacts with the web page, a trigger is activated and the FrontEnd collects the data, putting them into the integration data structure and sending them to the BackEnd 1. Using a WebSocket connection, the page subscribes to a specific topic that belongs to BackEnd 1, to be able to retrieve results from the backend asynchronously. After a run is launched, a pop-up confirms the successful job submission to the backend. 
After the algorithm has run, the output is stored into the browser cache using a predefined data model, together with all the needed metadata. This data model provides a common JSON structure to pass data from the UI to the BackEnd 1, and vice versa.
Now, it can be retrieved by the FrontEnd using a set of functions and the results can be used to display the proper page. 

\subsection{BackEnd 1}

As reported in the picture of the architecture, the BackEnd 1 is made of a single Java application hosted on IBM Cloud Foundry, which communicates with Cloudant~\cite{ibm:dant} to read configuration parameters needed to call the Cloud Functions. To be more specific, the data model is composed by just one collection called ‘config’ that contains the parameters that let BackEnd 1 to call the Cloud Functions. In the following we show an example of how the BackEnd 1 is linked to the database, this allows to retrieve data and to recall the cloud functions:

\begin{verbatim}

{
  "_id": "smile_super_position",
  "functionHttpMethod": "POST",
  "functionBackendUrl": "URL",
  "functionParams": {
    "body": "incomingRequestBody",
    "headers": {
      "Authorization": "IAMBearerToken",
      "Content-Type": "application/json",
      "Accept": "application/json"
    }
  }
}

\end{verbatim}
The BackEnd 1 reads these data and translates this configuration into a restful API call. All the details which make it possible are given: the method (POST), the endpoint, the authentication and the header. Given a new cloud function, the addition of new functionalities to this asset is as simple as adding another record into the config collection. This way, the architecture guarantees extensibility and it can be easily reused in future integration.
Once the job is submitted to the quantum computer via the quantum API provided in this case by Qiskit, an attribute that contains a random Event Streams topic (e.g.: topic-1234) is read from the queue. User segregation is achieved since each client will be listening on a single specific topic. Concerning the development, we adopted Java Spring Boot\cite{webb2013spring} as a framework to enable a “production-ready" environment benefiting from the automatic Spring configuration and third-party libraries management. 

\subsection{Cloud Functions}
The serverless architectures allow the developers to focus on business logic exclusively without worrying about preparing the runtime, managing deployment and infrastructure related concerns, in this case a quantum computing integration~\cite{8939081}.
The Cloud Functions component~\cite{ibm:cf} defined in the proposed framework is the one responsible to build the quantum algorithm and execute the related circuit on the IBM Quantum provider, giving the possibility to choose between the available quantum hardware or simulators.
From an architectural point of view, this is a block of instructions that runs on IBM Cloud Functions. This block of code, that is called “action”, is invoked by the BackEnd 1 to process a user request received from the front end layer.
The code of the action is written in Python, the language adopted by the Qiskit library.
When the action is called from the BackEnd 1, a quantum circuit is created according to the input parameters. Then, an IBM Quantum job is defined and it is sent to the IBM Quantum system for the execution.
Serverless architectures are gaining traction in cloud-based application architectures used by startups and matured organizations alike~\cite{8529465}. 

Instead of waiting for its final results, we collect its job ID\footnote{\url{https://qiskit.org/documentation/_modules/qiskit/providers/ibmq/job/ibmqjob.html}} as a parameter that is provided to the BackEnd 2 on a Event Streams queue.

From an application point of view, the action is invoked by BackEnd 1 with the following set of parameters: 
\begin{itemize}
    \item algorithm-specific input parameters needed to build the related quantum circuit;
    \item the type of IBM Quantum backend requested: real quantum hardware, noiseless simulator or noisy simulator;
    \item the client ID and process job ID\footnote{Not the IBM Quantum job ID.}; these two parameters are simply transmitted to BackEnd 2 and they are not used by the action.
\end{itemize}

Once the afore mentioned parameters are acquired, the action creates the corresponding quantum circuit. The action is authenticated via the quantum API used, in this case the IBM Quantum provider, and it runs the execution request on the selected quantum backend.

When the algorithm execution has been selected to run on a real quantum device, the code selects automatically the least busy IBM Quantum backend according to the number of qubits needed to map the quantum circuit. 
When the selection is the quantum simulator, the  backend is set to be the “qasm simulator".
This option is used both in case of a noiseless simulation and in case of a noisy simulation, where the action uses the "qasm simulator" activating the noise model of the least busy real device.

After the execution command, the action sends to the Event Streams queue the IBM Quantum provider job ID, the selected quantum backend, and the client and job ID received by the BackEnd 1. In case of error during the job submission to the IBM Quantum provider, or during the transmission of the parameters to the Event Streams queue, the action returns the associated error message.

\subsection{BackEnd 2}

The BackEnd 2 is a contenierized application whose main task is to retrieve results of a quantum job as soon as they are available, regardless of the selected IBM Quantum backend, that could be anyone of the available quantum devices or cloud simulators.

BackEnd 2 activity focuses on the management of the job ID polling, and this is performed by the integration of two framework components:

\begin{itemize}
    \item Kafka client: this component enables communication between the BackEnd 1, the BackEnd 2 and the Cloud Functions component (both in input and in output);
    \item Polling: this component uses Qiskit libraries to retrieve a job final result and it uses the Event Stream component to deliver results back to BackEnd 1.
\end{itemize}

The BackEnd 2 is written and deployed on a Python IBM Cloud Foundry instance. In addition, it leverages Python multi-process module to enable the management of more than one request at a time \footnote{\url{https://docs.python.org/3.7/library/multiprocessing.html}}.
\begin{figure}[h]
\centering
\includegraphics[width=0.45\textwidth]{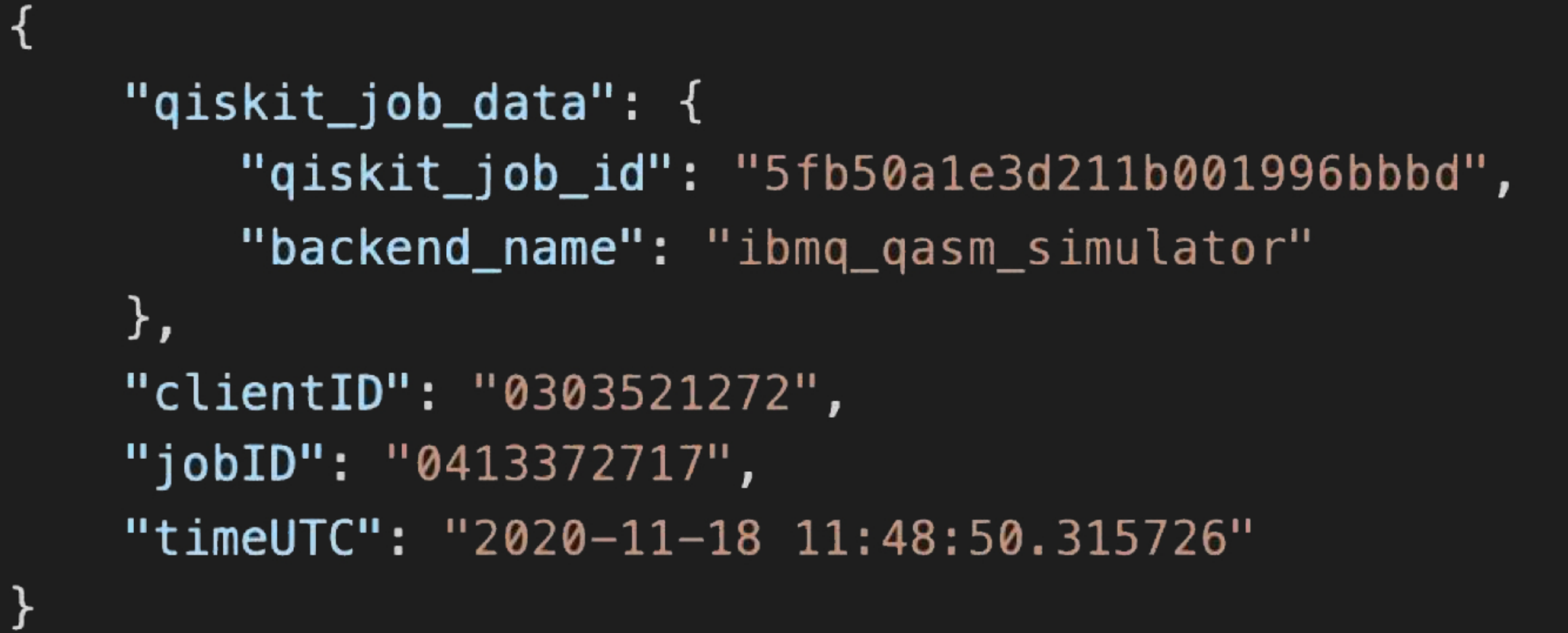}
\caption{Example of input JSON.}
\label{img:json1}
\end{figure}
The main function is single-process and it continuously gets events from the IBM Event Streams input topic, putting them into a manager pool queue (process-safe) that is shared between the running processes. The main function also initiates the pool workers, triggered by an event that consists of a JSON sent to the topic by the Cloud Functions component, and that contains all the relevant information about the job. 
In Fig. \ref{img:json1} we show an example of an event's input JSON.

\begin{figure*}[ht]
\centering
\includegraphics[width=.9\textwidth]{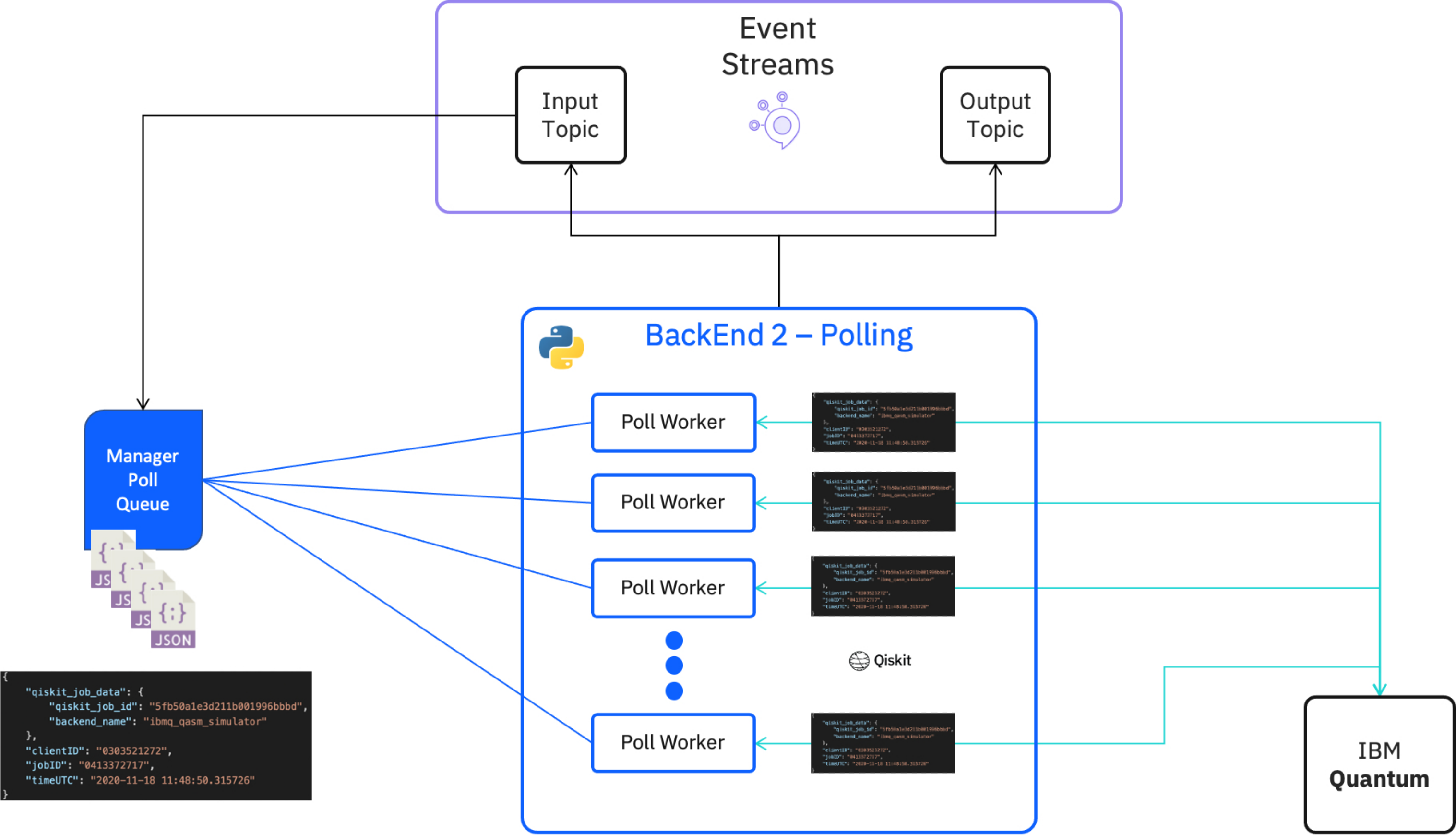}
\caption{Polling mechanism overview.}
\label{img:be2}
\end{figure*}

In the Polling component, each independent process worker gets a job from the shared queue, and monitor it until the result is ready. In Fig. \ref{img:be2} is reported the overview of the Polling mechanism inside the BackEnd 2.

\subsubsection{Kafka Client Component}
The Event Streams component consists of two Python classes that implement connections with the input and output components of the queue. They lie on the same Event Streams instance managing different topics.
The Kafka-based Event Streams component allows to realize a smart decoupling between the quantum and classical world enabling asynchronous communication mechanisms needed to reduce the connection time to the quantum on-line system.
The input component is represented by the insertion from the Cloud functions of the JSON containing the call execution id and the related job ID. They were produced submitting a quantum circuit to be executed on the IBM Quantum system. This part is implemented by a class that interacts with the queue in two ways:
\begin{itemize}
    \item listen as “consumer” to retrieve the incoming messages and then activate the polling activity through the “Polling component”;
    \item write as “producer” to re-insert in the queue the unprocessed job ID. It happens when, considering the time between the “estimated time” of the process and the actual timestamp exceeds a pre-defined threshold. This control has been developed to reduce as much as possible the polling time on the IBM Quantum system.
\end{itemize}
The output component of the queue is where the BackEnd 2 writes the results retrieved from IBM Quantum. The component then makes the results available to the BackEnd 1, that will proceed to post-process and to retrieve them back to the front end.
The class is designed to write as “producer” on the output queue and to send JSON extended input together with the retrieved results or the error messages.

\subsubsection{Polling Component}
This component contains all the dependencies related to the Qiskit framework.
The task of the Polling component is to get, using the Event Streams component, an IBM Quantum job launched by the Cloud Functions on BackEnd 1 proper input, monitoring its progress and returning the results to the BackEnd 1.
This component is initialized in the main function as poll workers: each worker gets an event from the manager pool queue, processing it independently from each others.

It consists of three modules:

\begin{itemize}
    \item PollingIBMQWorker: contains the worker started in the main function;
    \item PollingIBMQ: a class, whose method called “retrieveAndDeliverResult", contains all the logic of the process for retrieve and manage each job's result from IBM Quantum;
    \item QuantumUtils: contains functions for the authentication to the IBM Quantum cloud.
When each worker is started, the IBM Quantum account is enabled using its API key authentication, and it creates the connection to the IBM Quantum systems.
\end{itemize}

As a first step, the PollingIBMQ class checks if the job exists on the quantum backend and then checks its final state (e.g. DONE, CANCELED or ERROR). If the job is in one of the allowed final states, the PollingIBMQ sends it to the output topic, and in case the final status is DONE it adds the obtained results. 

If the job is not complete and if the quantum backend is a real quantum device, it checks what is its estimated completion time\footnote{provided by the queue\_info() function (\url{https://qiskit.org/documentation/\_modules/qiskit/providers/ibmq/job/ibmqjob.html##IBMQJob.queue\_info})}: if it is less than a defined time, it will wait for the job to be finished using Qiskit proper function; otherwise, it will send the event back to the input topic so that it can process another job. In the case of a quantum simulator backend, it automatically starts waiting for the job, because the estimated time is not provided by Qiskit as waiting time is usually much less.
The estimated completion time is then added to the JSON, so that it can be used by the Event Streams component to filter the event. 
If the job ends in a final error state or if there is an error in the process, this component always returns a result with the description of the encountered error.

This strategy gives priorities to the jobs based on results readiness, to better manage multiple requests and, likewise, reducing the waiting time for the frontend users.

\begin{figure}[h]
\centering
\includegraphics[width=0.45\textwidth]{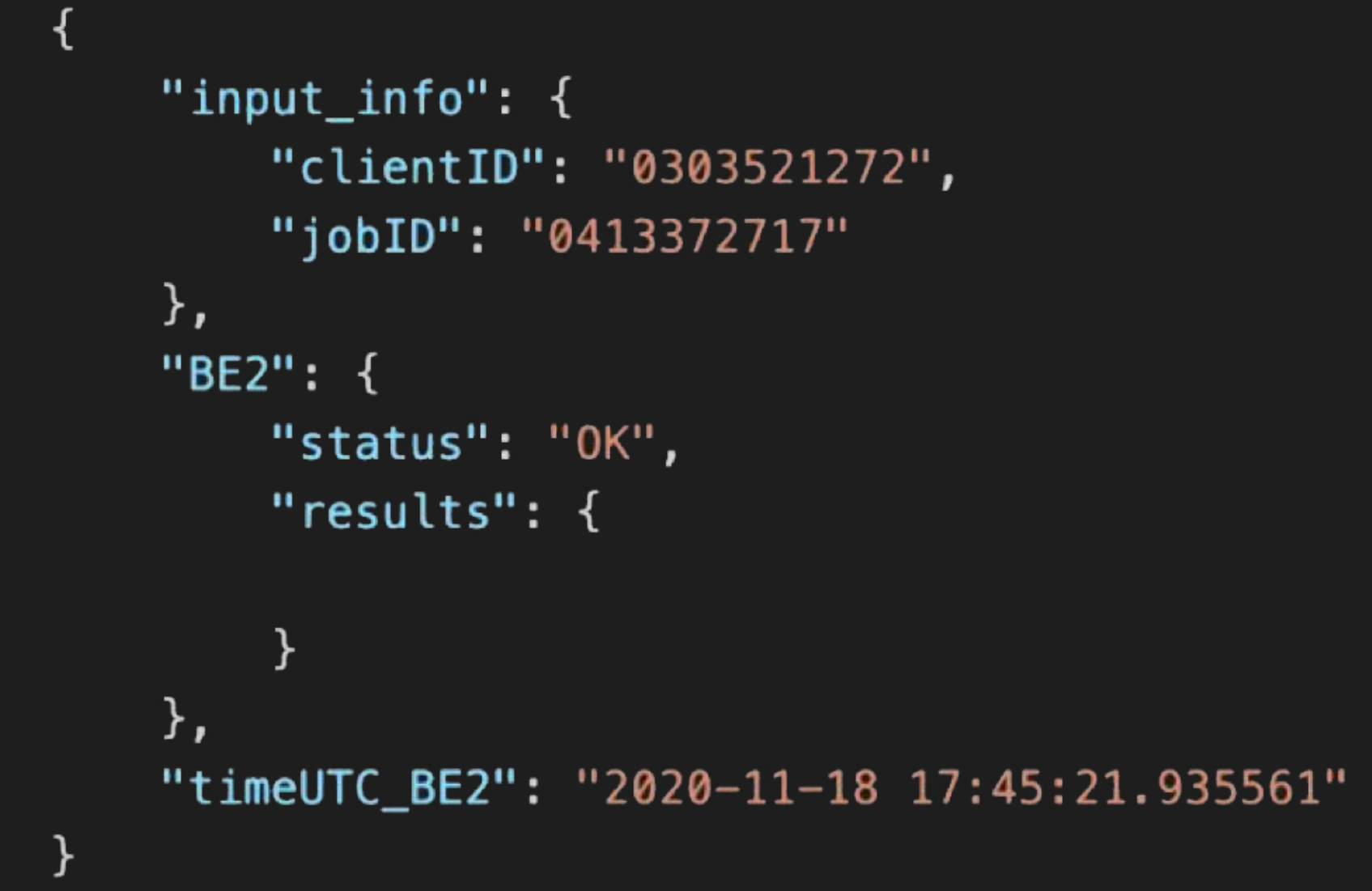}
\caption{Example of final JSON.}
\label{img:finaljson}
\end{figure}

In Fig. \ref{img:finaljson} there is an example of the final JSON.

\section{Conclusions}
\label{sec:concl}
The potential of an innovative technology such as quantum computing is evident in the number of scientific articles that show its advances. The real scope of the technology and its adoption depend on the possibility and practicality of integrating it into the current context. 
To facilitate this adoption, in this paper we define the criteria and critical issues for a possible integration of quantum computing within a cloud computing architecture.
After an introduction about a general architecture and its operation using various cloud computing technologies, we even describe a real implementation via a Minimum Viable Product. In this context we developed a scheduler that allows users to correctly schedule and manage several quantum jobs. This is a simple solution that can be expanded and it can be downloaded from GitHub. In this scenario we used the IBM Quantum API tool, however its role can be extended without loss of generalization to any other Quantum API tools. The proposed solution can be schematically described in the following:\\
\subsubsection*{Why}
\begin{itemize}
    \item nowadays, quantum algorithms can be run on different web platforms, where the user needs to write an algorithm using the available tools;
    \item programmatic access via API can be done using Software Development Kits (SDKs);
    \item As the interest in quantum computing continues to grow, it is urgent to find methods to fill the gap between a low-level approach and the high-level general user experience.
\end{itemize}

\subsubsection*{What}
The proposed framework provides developers, UI designers and researchers with a system that:

    \begin{itemize}
        \item enables and speeds up the creation and the deployment of web applications with a hybrid classical-quantum backend;
        \item creates a custom user experience based on the problem to be solved;
        \item spreads out the usage of quantum computing on Customer's Production environments.
    \end{itemize}

\subsubsection*{How}

 \begin{itemize}
 \item 
This method has been reflected in a well-defined framework built as MVP on IBM Cloud and Quantum technologies:
    \begin{itemize}
        \item IBM Cloud Foundry;
        \item IBM Cloudant;
        \item IBM Event Streams;
        \item IBM Cloud Functions;
        \item IBM Quantum.
    \end{itemize}
\end{itemize}
\begin{itemize}
    \item A new approach on asynchronous job submission method specially created to support hybrid classical- quantum web applications.
    \item Best practices: configuration driven architecture, open source, loosely coupled architectural pattern to schedule computation (polling/batch).
\end{itemize}

\paragraph{Acknowledgements}
We would like to acknowledge G. De Sio and L. Savorana for grateful discussion and interaction. We acknowledge use of the IBM Quantum for this work. The views expressed are those of the authors and do not reflect the official policy or position of IBM or the IBM Quantum team. IBM, the IBM logo, and ibm.com are trademarks of International Business Machines Corp., registered in many jurisdictions worldwide. Other product and service names might be trademarks of IBM or other companies. The current list of IBM trademarks is available at \url{https://www.ibm.com/legal/copytrade}.

\section{References}

\printcredits

\bibliographystyle{cas-model2-names}

\bibliography{cas-refs}

\vskip3pt

\end{document}